# Covariant Energy-Momentum Conservation In General Relativity

## *by Philip E. Gibbs* [1]

Abstract

A covariant formula for conserved currents of energy, momentum and angular-momentum is derived from a general form of Noether's theorem applied directly to the Einstein-Hilbert action of classical general relativity. Energy conservation in a closed big-bang cosmology is discussed as a special case. Special care is taken to distinguish between kinematic and dynamic expressions.

[1] e-mail: philip.gibbs@pobox.com
27 Stanley Mead, Bristol, BS12 0EF, UK

**Is Energy Conserved in General Relativity?**

This may seem a strange question to be asking in 1997 considering that conservation of energy in general relativity was established by Einstein himself in 1916, the year after the theory was published [1]. Yet there remains a sour taste to Einstein's pseudo-tensor formulation because of its lack of covariance. Modern attempts to improve on it can only really be said to be successful in specific cases such as asymptotically flat space-times and different forms of pseudo-tensor have only added to the ambiguity.

To exemplify the problem it is helpful to ask again a question which is often posed by students. *Where does the energy of the Cosmic Background Radiation go when it is red-shifted?* Each photon carries a quantum of energy which slowly decreases as the universe expands. Ignoring interactions with matter which are irrelevant on the cosmological scale, the total number of photons remains constant. The total energy in the radiation must therefore decrease, where does it go? To avoid objections let us assume that the universe is closed so that the total number of photons is finite.

The obvious answer is that the energy is transferred to the gravitational field which has an increasing (negative) total energy. But to justify this response we must be able to provide a general formula for energy which can be applied to this case. The pseudo-tensor is not sufficient because it is co-ordinate dependent and it is not possible to define a co-ordinate system over a whole spatial hypersurface of the closed universe if its topology is that of a three-sphere. How can we then be sure that we have not missed a leak of energy at the inevitable co-ordinate singularity where the formulation breaks down?

It may well be possible to stop the gap with some mathematical slog, but a survey of most of the popular textbooks on general relativity does not turn up such an analysis. Those which use modern differential geometry shy away from the pseudo-tensor altogether and often leave general energy-momentum conservation outside the scope of the book. This is surprising because there *is* an elementary covariant way to treat energy and momentum conservation as I will show in this paper. Aside from the example I have given, the problem is important because of other cases such as energy-momentum transferred by gravitational waves and the contribution of gravitational self energy to mass which has been verified in tests of the strong equivalence principle by laser ranging of the moon.

An interesting twist to the problem of total energy in a closed universe is that it comes out to be exactly zero rather than an arbitrary constant. This fact is often quoted when people ask where all the energy came from to set off the big bang, the answer: There is none! Again, it would be useful to have a tidy way of proving this. Sometimes the null total energy is taken as an indication that energy conservation in general relativity is an empty statement. This is not true because the energy is only zero dynamically, not kinematically. I.e. it is only zero when the gravitational field equations are satisfied. To see this we must provide the right formula for energy which can be applied to kinematics.

**Energy and Noether's Theorem**

Let us turn now to the gravitational field equations,

$$R^{\mu\nu} - \tfrac{1}{2} R g^{\mu\nu} = 8\pi G T^{\mu\nu}$$

By the Bianchi Identities, the divergence of the left hand side vanishes and implies,

$$T^{\mu\nu}{}_{;\nu} = 0$$

If we were doing special relativity the divergence of the energy-momentum stress tensor would be adequate to establish energy-momentum conservation, but in general relativity it is not unless there is a killing vector field $K^\alpha$, which generates an isometry of the metric,

$$\delta x^\alpha = \varepsilon K^\alpha$$
$$\Delta g_{\alpha\beta} = \varepsilon(K_{\alpha;\beta} + K_{\beta;\alpha}) = 0$$
$$J^\mu = K^\alpha T_\alpha{}^\mu \Rightarrow J^\mu{}_{;\mu} = 0$$

When we have such a divergenceless *vector* current for energy flow, we can derive conservation because,

$$J^\nu{}_{;\nu} = \frac{1}{\sqrt{g}} \frac{\partial \sqrt{g} J^\nu}{\partial x^\nu}$$

and therefore the current can be integrated over a closed space-like hypersurface of constant time co-ordinate to give a conserved Energy,

$$E = \int \sqrt{g} J^0 d^3 x$$

When we have no Killing field and we try this with the energy-momentum tensor we get,

$$\sqrt{g} T^\nu{}_{\mu;\nu} = \frac{\partial \sqrt{g} T^\nu{}_\mu}{\partial x^\nu} - \tfrac{1}{2} \sqrt{g} T^{\lambda\nu} \frac{\partial g_{\lambda\nu}}{\partial x^\mu} = 0$$

The second term, which can not be integrated, signifies a flow of energy between matter and the gravitational field. A term representing a contribution to the energy momentum tensor from the gravitational field is required. In deriving it Albert Einstein got some help from Emmy Noether who provided a general theorem relating conservation laws to symmetries of Lagrangians [2]. Noether's theorem could be applied to the Hilbert-Einstein action for gravitation after an integration by parts designed to remove second derivatives of the metric tensor from the action. This leads to Einstein's non-covariant energy-momentum pseudo-tensor for which,

$$\frac{\partial}{\partial x^\mu}(\sqrt{g}T^\mu{}_\nu + t^\mu{}_\nu) = 0$$

A concise exposition of the derivation can be found in Dirac's book on General Relativity [3]. After further manipulations which are described in the book of Landau and Lifshitz [4], a version of the pseudo-tensor which is symmetric in the two indices can be found. This symmetry is important if conservation of angular momentum is also to be demonstrated, but it makes the analysis even more complicated and less well motivated since Noether's theorem is not used directly. Many other forms of the pseudo-tensor are possible such as the symmetric one given by Weinberg [5]. Recently, Bak, Cangemi and Jackiw derived a symmetric pseudo-tensor using Noether's theorem and provided a useful comparison with other pseudo tensors [6]. Their result is still not covariant and is presented in a form which is divergenceless even kinematically. The full formula for the energy which shows the contribution to energy from different sources, and which is conserved dynamically but not kinetically, is often simplified using the field equations.

With so many different versions of energy-momentum conservation which do not give exactly the same answer, one might ask what energy really is. The different formulations all differ by a quantity which is kinematically conserved (perhaps even before the simplification using the field equations) and which is asymptotically negligible in the weak field approximation to gravitational waves. Even so, we might have expected a unique concept of energy.

The correct answer to the question "What is energy?" is that the correct energy is that which is given by Noether's theorem. This is still not unique since the Lagrangian is only unique up to a term which is a perfect divergence. This feature is what is used to get rid of the terms with second derivatives in the metric tensor before deriving the energy equation, but it also destroys covariance. Even with covariance other forms of the action are known but at least it might be worth while having the correct expression for energy for the traditional Einstein-Hilbert action.

$$S = \frac{1}{16\pi G}\int(-\sqrt{g}R + L_M)d^4x$$

This can be done using a form of Noether's theorem which works with second derivatives in the Lagrangian.

## Energy Momentum in GR

Consider a general action of the form,

$$S = \int L(\phi^a, \phi^a{}_{,\mu}, \phi^a{}_{,\mu\nu}) d^4x$$

$\phi^a$ are all the components of the field variables including the metric tensor. Applying a variation principle, the Euler-Lagrange equations are,

$$\frac{\partial L}{\partial \phi^a} - \partial_\mu \left( \frac{\partial L}{\partial \phi^a{}_{,\mu}} \right) + \partial_{\mu\nu} \left( \frac{\partial L}{\partial \phi^a{}_{,\mu\nu}} \right) = 0$$

A continuous symmetry of the action which might include co-ordinate changes as well as field transformations would have a general infinitesimal form,

$$\delta \phi^a = \varepsilon \theta^a$$

$$\delta x^\mu = \varepsilon k^\mu$$

and invariance implies that,

$$\delta L = \frac{\partial L}{\partial \phi^a} \varepsilon \theta^a + \frac{\partial L}{\partial \phi^a{}_{,\mu}} \varepsilon \theta^a{}_{,\mu} + \frac{\partial L}{\partial \phi^a{}_{,\nu\mu}} \varepsilon \theta^a{}_{,\mu\nu} = \partial_\mu (\varepsilon k^\mu L)$$

Combining results we finally derive the conserved Noether current,

$$I^\mu = \frac{\partial L}{\partial \phi^a{}_{,\mu}} \theta^a - \partial_\nu \left( \frac{\partial L}{\partial \phi^a{}_{,\mu\nu}} \right) \theta^a + \frac{\partial L}{\partial \phi^a{}_{,\nu\mu}} \theta^a{}_{,\nu} - k^\mu L$$

$$\Rightarrow I^\mu{}_{,\mu} = 0$$

$$J^\mu = \tfrac{1}{\sqrt{g}} I^\mu$$

$$\Rightarrow J^\mu{}_{;\mu} = 0$$

## The Covariant Formulation

The next step is to apply this form of Noether's theorem directly to the Einstein-Hilbert action. The action is invariant under diffeomorphisms (co-ordinate transformations). Infinitesimal diffeomorphisms are generated by

$$\delta x^\mu = \varepsilon k^\mu$$
$$\delta g_{\alpha\beta} = \varepsilon \Delta g_{\alpha\beta} = \varepsilon(k_{\alpha;\beta} + k_{\beta;\alpha})$$
$$\psi^i = \varepsilon \theta^i(k^\mu)$$

$\psi^i$ are all the field components other than the metric tensor. The answer will not take the form of an energy-momentum tensor, it will be a contravariant vector current which depends on the contravariant transport vector field $k^\alpha$ and by Noether's theorem it will be conserved for any choice of $k^\alpha$. This is a natural consequence of the fact that in general relativity the symmetry is diffeomorphism invariance rather than global Lorentz invariance.

The derivation is quite long but will nevertheless by given in some detail for the gravitational part. Our starting point is the Lagrangian density,

$$L = \tfrac{-1}{16\pi G}\sqrt{g}\,R + L_M$$

It is assumed that the matter Lagrangian contains no *higher than first* covariant derivatives in field variables, and transforms as a scalar-density under diffeomorphisms In that case the matter field contribution to Noether current is given by,

$$J^\mu = J^\mu{}_G + J^\mu{}_M$$
$$J^\mu{}_M = k^\nu T_\nu{}^\mu$$

The Riemann curvature tensor is given by,

$$R_{\mu\nu\rho\sigma} = R^{(1)}{}_{\mu\nu\rho\sigma} + R^{(2)}{}_{\mu\nu\rho\sigma}$$
$$R^{(1)}{}_{\mu\nu\rho\sigma} = \tfrac{1}{2}(g_{\mu\sigma,\nu\rho} - g_{\nu\sigma,\mu\rho} - g_{\mu\rho,\nu\sigma} + g_{\nu\rho,\mu\sigma})$$
$$R^{(2)}{}_{\mu\nu\rho\sigma} = \Gamma_{\beta\mu\sigma}\Gamma_{\alpha\nu\rho}g^{\alpha\beta} - \Gamma_{\beta\mu\rho}\Gamma_{\alpha\nu\sigma}g^{\alpha\beta}$$
$$\Gamma_{\mu\nu\sigma} = \tfrac{1}{2}(g_{\mu\nu,\sigma} + g_{\mu\sigma,\nu} - g_{\nu\sigma,\mu})$$
$$R = R_{\mu\nu\rho\sigma}g^{\mu\sigma}g^{\nu\rho}$$

The gravitational contribution to the current will be split up accordingly. The factor $\sqrt{g}$ passes outside the partial derivatives but an extra term must be included when it passes through the co-ordinate derivatives.

$$J^{\mu}{}_{G} = \tfrac{1}{16\pi G}(k^{\mu} R - j^{\mu})$$

$$j^{\mu} = \left\{\frac{\partial R^{(2)}}{\partial g_{\alpha\beta,\mu}} - (\partial_{\nu} + \Gamma^{\sigma}{}_{\nu\sigma})\left(\frac{\partial R^{(1)}}{\partial g_{\alpha\beta,\mu\nu}}\right)\right\}\Delta g_{\alpha\beta} + \frac{\partial R^{(1)}}{\partial g_{\alpha\beta,\mu\nu}}(\Delta g_{\alpha\beta})_{,\nu}$$

$$\Delta g_{\alpha\beta} = k_{\alpha;\beta} + k_{\beta;\alpha}$$

and we can calculate,

$$\frac{\partial R^{(1)}}{\partial g_{\alpha\beta,\mu\nu}} = g^{\alpha\beta}g^{\mu\nu} - g^{\mu(\beta}g^{\alpha)\nu}$$

If we knew that the answer must be covariant, then we could use the fact that, in a local inertial frame, the first derivatives of the metric tensor vanish at a single point. This makes all the difficult terms in the current vanish and the calculation is easy to complete.

$$j^{\mu} = (g^{\alpha\beta}g^{\mu\nu} - g^{\mu(\beta}g^{\alpha)\nu})(\Delta g_{\alpha\beta})_{;\nu}$$

However, covariance has not yet been proven, so the terms should be worked out at length to be sure of the result.

$$R^{(2)} = \Gamma_{\sigma\tau}{}^{\tau}\Gamma^{\sigma}{}_{\nu}{}^{\nu} - \Gamma_{\sigma\tau\nu}\Gamma^{\sigma\tau\nu}$$

$$\frac{\partial R^{(2)}}{\partial g_{\alpha\beta,\mu}} = g^{\beta\mu}\Gamma^{\alpha}{}_{\nu}{}^{\nu} + g^{\alpha\mu}\Gamma^{\beta}{}_{\nu}{}^{\nu} - g^{\alpha\beta}\Gamma^{\mu}{}_{\nu}{}^{\nu} - \Gamma^{\alpha\beta\mu} - \Gamma^{\beta\alpha\mu} + \Gamma^{\mu\alpha\beta}$$

$$g^{\alpha\beta}{}_{,\nu} = -\Gamma^{\alpha\beta}{}_{\nu} - \Gamma^{\beta\alpha}{}_{\nu}$$

$$\partial_{\nu}\left(\frac{\partial R^{(1)}}{\partial g_{\alpha\beta,\mu\nu}}\right) = -\tfrac{1}{2}\Gamma^{\alpha\beta\mu} - \tfrac{1}{2}\Gamma^{\beta\alpha\mu} - g^{\alpha\beta}(\Gamma^{\mu\nu}{}_{\nu} + \Gamma^{\nu\mu}{}_{\nu}) +$$

$$\Gamma^{\mu\beta\alpha} + \tfrac{1}{2}g^{\mu\beta}(\Gamma^{\alpha\nu}{}_{\nu} + \Gamma^{\nu\alpha}{}_{\nu}) + \tfrac{1}{2}g^{\mu\alpha}(\Gamma^{\beta\nu}{}_{\nu} + \Gamma^{\nu\beta}{}_{\nu})$$

$$j^{\mu} = \{g^{\beta\mu}(\Gamma^{\alpha}{}_{\nu}{}^{\nu} - \Gamma^{\nu\alpha}{}_{\nu}) + g^{\alpha\beta}\Gamma^{\nu\mu}{}_{\nu} - \Gamma^{\alpha\beta\mu}\}\Delta g_{\alpha\beta} +$$

$$\{g^{\alpha\beta}g^{\mu\nu} - g^{\mu\beta}g^{\nu\alpha}\}(\Delta g_{\alpha\beta,\nu} - \Gamma^{\sigma}{}_{\nu\sigma}\Delta g_{\alpha\beta})$$

$$= \{g^{\alpha\beta}g^{\mu\nu} - g^{\mu\beta}g^{\nu\alpha}\}\Delta g_{\alpha\beta;\nu}$$

The current can be expressed in terms of the transport vector field,

$$J^{\mu}{}_{G} = \tfrac{1}{16\pi G}(k^{\mu}R - 2k^{\alpha}{}_{;\alpha}{}^{\mu} + k^{\alpha;\mu}{}_{\alpha} + k^{\mu;\alpha}{}_{\alpha})$$

This is the final covariant expression for the contribution to energy-momentum current from the gravitational field. The dependency on the contravariant transport vector $k^{\alpha}$ is left explicit. Any attempt to remove it to derive a stress tensor will destroy covariance and make the result co-ordinate dependent. The generality of the expression would be lost. The term with the curvature scalar depends on second derivatives of the metric field in addition to first derivatives. This is unusual for an expression of energy or momentum. The second derivatives could be removed by adding a term which is kinematically divergence free but only at the cost of covariance.

The choice of transport vector field is used to distinguish between currents of energy, momentum, angular-momentum, etc. In the case where it is everywhere time-like the current is interpreted as flow of energy. Energy and momentum are therefore not unique but this is no surprise or obstacle. For a given co-ordinate system the energy and momentum which are conjugate to time and space co-ordinates can be found but they do not form a tensor.

Using the identity,

$$k^{\alpha}{}_{;\alpha\mu} - k^{\alpha}{}_{;\mu\alpha} = k^{\alpha}R_{\alpha\mu}$$

and including the contribution from the matter fields we get

Energy Momentum in GR

$$J^\mu = \tfrac{1}{16\pi G}(k^\mu R - k^{\alpha\ ;\mu}_{\ ;\alpha} + k^{\mu\ ;\alpha}_{\ ;\alpha} - 2k^\alpha R_\alpha^{\ \mu}) + k^\alpha T_\alpha^{\ \mu}$$

If we now apply the gravitational field equations the expression can be simplified dramatically.

$$J^\mu = \tfrac{1}{16\pi G}(k^{\mu\ ;\alpha}_{\ ;\alpha} - k^{\alpha\ ;\mu}_{\ ;\alpha})$$

This can also be written in a form without explicitly covariant derivatives,

$$J^\mu = \tfrac{1}{16\pi G \sqrt{g}}\left[\sqrt{g}(k_{\nu,\beta} - k_{\beta,\nu})g^{\mu\nu}g^{\alpha\beta}\right]_{,\alpha}$$

The divergence of this expression is identically zero but it is only correct dynamically. The full expressions we derived before simplifying with the field equations are kinematic formulae for the energy-momentum currents and their divergence is not identically zero, but it is dynamically zero. They show directly the contributions to energy and momentum from separate parts of the Lagrangian.

These final dynamic expressions are the Komar superpotential [7]. When applied to an asymptotically flat space-time with appropriate use of boundary conditions this leads on to the Geroch-Winicour linkage formulation [8], the Bondi Energy and the ADM Energy [9].

**Energy in a Closed Cosmology**

We are now in a position to consider the case of energy in a closed cosmology. The covariant currents we have constructed are dynamically divergenceless. When integrated over a closed space-like hypersurface to give the total energy or momentum of the universe, the result must be zero because there is no boundary. As we noted in the introduction, this is hard to prove rigorously with a pseudo-tensor formulation, but with covariant currents it is a simple consequence of Gauss' theorem. In a closed universe with 3-sphere spatial topology there is no clear distinction globally between momentum and angular-momentum. They are consequences of rotations of the spatial 3-sphere about different points. The Minkowski model of energy-momentum 4-vectors and angular-momentum tensor is therefore not relevant to a closed spherical spatial topology.

For energy we must be careful about our choice of the transport vector field $k^\alpha$. If its covariant form $k_\alpha$ had components which were constant in some co-ordinate system then the current would be zero everywhere. That would not necessarily be incorrect but it should normally be chosen so that it increments the time co-ordinate uniformly. It follows that the energy current depends critically on the choice of time co-ordinate as is to be expected.

In a homogeneous Friedmann-Robertson-Walker cosmology there is a natural choice of time co-ordinate, that of the co-moving co-ordinate system. In a more realistic locally inhomogeneous cosmology the matter is not so simple. We could define cosmological time at each space-time event to be the maximum proper-time integrated along all time-like paths from the big-bang singularity to the event. This is well defined provided the cosmology has no closed time-like curves and can be sequentially foliated into space-like surfaces. It is easy to show that the surfaces of equal cosmological time are continuous space-like surfaces, except where they meet singularities which have a finite age. Even without singularities such surfaces are not smooth because co-ordinate systems which incorporate cosmological time are synchronous reference frames. It is known that they tend to develop co-ordinate singularities in realistic space-times (see [4]). In general, the energy currents can be defined relative to a smooth space-like foliation of space-time and a time-like transport vector field  It is beyond the scope of this article to establish when they exist.

Black-hole singularities do not actually cause a breakdown of energy-momentum conservation because it is possible to pass a space-like surface through a black hole which avoids the singularities. This can be done even at late times. The surface would have to bend back in time at nearly the speed of light  but can narrowly avoid becoming time-like. This can be seen in a Carter-Penrose diagram of a black-hole. Fortunately the total value of energy momentum and angular momentum for an isolated object depends *dynamically* only on the choice of transport vector at the boundary of the surface and not on the surface or choice of transport vector internally. This is not true *kinematically*.

Because of the difficulty of choosing a time co-ordinate which can be applied universally and which avoids all singularities, the interpretation of total energy in a closed universe is not perfectly clear. This is not too surprising or worrying since energy and momentum are usually considered to be relative, and the whole universe cannot be considered relative to anything else. Nevertheless, there *are* covariantly conserved energy and momentum currents whose total integrates to zero. This is not a trivial result since it is not true kinematically if we use the correct kinematic definition of the currents which is derived directly from Noether's theorem *without* using the field equations to simplify. Remember that it is only this kinematic form which shows explicitly the contribution to the currents from individual fields.

**Conclusions**

I have shown how to derive conserved energy and momentum currents by applying Noether's theorem directly to the Hilbert-Einstein action of general relativity. I have not entered into questions of lacalisability of the gravitational energy and have not considered how to choose the transport vector or boundary conditions to obtain physically useful answers in particular cases except for a closed cosmology.

**Note Added**

I am informed that the Komar superpotential [7] may have been derived from a variational principle in this way by Trautman in a conference paper in 1961. A more detailed derivation was given by Kijowski in 1978 [11]. It is my pleasure to thank Piotr Chrusciel, Joseph Katz, Jerzy Kijowski, Greg Oganessyan, and Kumar Shwetketu Virbhadra for these and other useful references and comments.

Energy Momentum in GR